\PassOptionsToPackage{numbers,sort&compress}{natbib}  
\documentclass[preprintnumbers,amsmath,amssymb,prd,notitlepage,nofootinbib,superscriptaddress]{revtex4}

\makeatletter
\renewcommand{\p@subsection}{}
\renewcommand{\p@subsubsection}{}
\makeatother

\usepackage[numbers,sort&compress]{natbib}
\usepackage[normalem]{ulem}
\usepackage{graphicx}
\usepackage{xcolor}
\usepackage{dcolumn}
\usepackage{bm}
\usepackage{subfigure}
\usepackage{wasysym}
\usepackage{verbatim}
\usepackage{color}
\usepackage{amsmath}
\usepackage[utf8]{inputenc}
\usepackage[hidelinks]{hyperref}
\usepackage{aas_macros}

\newcommand{\rmn}{\mathrm}
\newcommand{\N}{\mathcal{N}}

\newcommand{\balpha}{\boldsymbol{\alpha}}

\newcommand{\Planck}{{\it Planck}~}                                  

\newcommand{\be}{\begin{equation}}                                  
\newcommand{\ee}{\end{equation}}                                    
\newcommand{\ba}{\begin{eqnarray}}                                  
\newcommand{\ea}{\end{eqnarray}}                                    
\newcommand{\nn}{\nonumber}                                          
\newcommand{\barr}{\begin{array}}                                    
\newcommand{\earr}{\end{array}}                                      
                          
\newcommand{\vsp}{\vphantom{\Big[}\\}

\def\n{{\bf n}}

\def\hphi{{\hat\phi}}

\def\n{{\boldsymbol{n} }}
\def\l{{\boldsymbol{\ell} }}
\def\lo{{\boldsymbol{\ell}_1 }}
\def\lt{{\boldsymbol{\ell}_2 }}
\def\L{{\bf L}}
\def\tX{{\tilde X}}
\def\tY{{\tilde Y}}
\def\tU{{\tilde U}}
\def\hphi{{\hat\phi}}
\def\RDN0{\mbox{RDN}^{(0)}}
\def\Nz{N^{(0)}}
\def\None{N^{(1)}}
\def\Ntwo{N^{(2)}}
\def\N32{N^{(3/2)}}
\def\Nzx{N^{(0)\times}}

\newcommand{\dotfac}[1]{({\L} \cdot {\boldsymbol{\ell}}_{#1})}

\def\Fl{g_{\lo\lt}}
\def\Fml{g_{-\lt,-\lo}}
\def\Fmlr{g_{-\lo,-\lt}}
\def\intlolt{\int_{\lo+\lt=\L}}

\begin{document}

\title{CMB lensing power spectrum estimation without instrument noise bias}

\author{Mathew S. Madhavacheril}
\email{mmadhavacheril@perimeterinstitute.ca}
\affiliation{Perimeter Institute for Theoretical Physics, \\ 31 Caroline Street N, Waterloo ON N2L 2Y5 Canada}
\author{Kendrick M. Smith}
\affiliation{Perimeter Institute for Theoretical Physics, \\ 31 Caroline Street N, Waterloo ON N2L 2Y5 Canada}
\author{Blake D. Sherwin}
\affiliation{Department of Applied Mathematics and Theoretical Physics, University of Cambridge,\\ Cambridge CB3 0WA, UK}
\affiliation{Kavli Institute for Cosmology, University of Cambridge,\\ Cambridge CB3 0HA, UK}
\author{Sigurd Naess}
\affiliation{Center for Computational Astrophysics, \\ Flatiron Institute, New York, NY, USA 10010}

\begin{abstract}
The power spectrum of cosmic microwave background (CMB) lensing will be measured to sub-percent precision with upcoming surveys, enabling tight constraints on the sum of neutrino masses and other cosmological parameters. Measuring the lensing power spectrum involves the estimation of the connected trispectrum of the four-point function of the CMB map, which requires the subtraction of a large Gaussian disconnected noise bias. This reconstruction noise bias receives contributions both from CMB and foreground fluctuations as well as instrument noise (both detector and atmospheric noise for ground-based surveys). The debiasing procedure therefore relies on the quality of simulations of the instrument noise which may be expensive or inaccurate. We propose a new estimator that makes use of at least four splits of the CMB maps with independent instrument noise. This estimator makes the CMB lensing power spectrum completely insensitive to any assumptions made in modeling or simulating the instrument noise. We show that this estimator, in many practical situations, leads to no substantial loss in signal-to-noise. We provide an efficient algorithm for its computation that scales with the number of splits $m$ as $\mathcal{O}(m^2)$ as opposed to a naive $\mathcal{O}(m^4)$ expectation. 
\end{abstract}

\maketitle
\flushbottom

\section{Introduction}
\label{sec:intro}

Over the past decade, the power spectrum of cosmic microwave background lensing has emerged as one of our most powerful probes of matter clustering. Building on first detections of the lensing power spectrum with the ACT and SPT experiments \cite{1103.2124,1202.0546,1105.0419} (which, in turn, built on the first detections of CMB lensing \cite{0705.3980,0801.0644}), measurements with the \Planck satellite have advanced the CMB lensing power spectrum to a mature and precise cosmological probe \cite{1303.5077,1502.01591,1807.06210}. However, with current- and next-generation ground-based CMB surveys, CMB lensing measurements and power spectra are expected to improve well beyond the current state of the art (e.g. \cite{1905.05777,2004.01139}). In particular, Simons Observatory \cite{1808.07445} and CMB-S4 \cite{1610.02743} have the potential to map lensing over a large-fraction of the sky and measure the lensing power spectrum at sub-percent precision for the first time. If such sub-percent-precision lensing power spectra can be robustly measured, they will yield a wealth of valuable information on fundamental physics, e.g. neutrino mass \cite{1509.07471}, primordial non-Gaussianity \cite{1710.09465} and dark matter properties \cite{1806.10165}. Encouraging progress has been made in addressing the major systematics concerns from astrophysical foregrounds \cite{1802.08230,1804.06403,1310.7547,2007.04325} and in theoretical modeling of the observables.

Despite the great promise of upcoming CMB lensing power spectrum measurements, crucial aspects of current methodology are likely inadequate for upcoming analyses with significantly higher signal-to-noise. A particular concern for upcoming ground-based CMB lensing measurements is that the complex and inhomogeneous noise structure arising from ground-based CMB experiments is difficult to simulate precisely; this, in turn, can make the important steps of $\Nz$ bias (or Gaussian bias) and mask-induced mean-field subtraction insufficiently accurate.

Bias subtraction is difficult to avoid in lensing reconstruction. Fundamentally, lensing power spectrum estimation measures the non-Gaussian part of the CMB four-point correlation function. The power spectrum of the lensing field estimator, since it is quadratic in the temperature or polarization fields, gives the full four-point correlation function. However, even in the absence of lensing, this four-point function has a non-zero expectation value. To isolate the non-Gaussian lensing signal, therefore, the disconnected or Gaussian part of the four-point function, commonly referred to as the $\Nz$ bias, must be subtracted off.

The $\Nz$ bias, which can be derived via Wick's theorem contractions of the four-point function, depends on both the CMB signal and noise power. A naive estimate of the $\Nz$ bias can be obtained by simple simulation. However, while CMB signal power is straightforward to model, the $\Nz$ bias depends also on the two-point correlator of the noise in the CMB map. Since the $\Nz$ bias is, on small but still relevant scales, typically larger than the signal by at least one order of magnitude, this implies that the noise power must be captured extremely accurately by the simulations. Otherwise, a mis-estimate of the $\Nz$ bias would result, which would cause a large error in our measurement of the lensing spectrum. The current state-of-the-art solution to this problem is the use of a realization dependent $\Nz$ algorithm, which estimates the $\Nz$ bias from a combination of data and simulations; while this absorbs small inaccuracies in the noise simulation, it breaks down if the noise simulations are very wrong,  and could be insufficiently accurate for future experiments. Another proposed solution is to avoid the $\Nz$ bias altogether by using different, disconnected regions of the Fourier plane to perform independent lensing reconstructions and then cross-correlating these reconstructions \cite{1011.4510}. However, this method is not immune to off-diagonal correlations in the instrument noise and requires iterated application to recover much of the signal-to-noise, which has not yet been demonstrated in practice.

Similar sensitivity to noise mis-simulation, especially on large scales, appears in the subtraction of the mean field, i.e. the response of the quadratic lensing estimator to non-lensing (noise, mask) statistical anisotropy. This is commonly estimated with simulations and subtracted off.

In this paper, we show how these problems can be avoided: the noise contribution to both $\Nz$ bias and mean field contribution can be entirely nulled by only using combinations of different splits of the data with independent noise. This `cross-only' method makes lensing power estimates insensitive to noise mis-simulation, with a minimal signal-to-noise penalty; even if the noise is modeled entirely incorrectly, the lensing power spectrum measurements will still be unbiased. 

An analogous procedure is already commonly used for estimating CMB temperature and polarization power spectra. These spectra are typically estimated using cross-spectra between data splits to avoid noise bias \cite{10.1086/377225}. However, naively, the combinatorics of performing an analogous procedure in the four-point correlation function are daunting; the algorithm for performing a cross-spectral estimation of the lensing power spectrum at only a modest numerical cost is a key result from our work. A naive combinatorics argument also suggests that we throw out a larger fraction of the data in a cross-only four-point estimator compared to a cross-only two-point power spectrum, but because CMB lensing typically uses scales that are signal-dominated over a large range, the signal-to-noise penalty often ends up being minimal even for the minimum of four splits.

For clarity, we define the term `instrument noise' that appears throughout this paper. This term refers to any contribution to power in observed CMB maps that is not nulled if two datasets taken far apart in time are differenced. Bolometer detector noise, readout noise and microwave emission from the atmosphere satisfy this property, since they are independent when separated by long intervals of time. It is possible that a contribution from microwave emission from the ground does not satisfy this property and ends up correlated between time-interleaved split maps. Such emission likely only contaminates the largest scales in the map; these are usually explicitly excluded or contribute negligible signal-to-noise in the lensing estimator. The lensed CMB and astrophysical foregrounds do not fall under the above definition; however their Gaussian contribution is very well understood from well motivated theoretical models and fits to millimeter-wavelength two-point  power spectra, and therefore will be adequately absorbed in the realization dependent noise bias subtraction (which we review in Section \ref{sec:review}) whether or not our cross-only estimator is used. The non-Gaussian signal from the lensed CMB is precisely the signal of interest we hope to isolate. Non-Gaussian bias from astrophysical foregrounds is well studied and promising methods exist for its mitigation \cite{1802.08230,1804.06403,1310.7547,2007.04325}.

This paper is structured as follows. In Section \ref{sec:review}, we introduce our notation
and briefly review CMB lensing reconstruction. We describe our new `cross-only' estimator and its fast implementation in Section \ref{sec:splits}. We conclude with a discussion in Section \ref{sec:discussion}. Throughout, we use a $\Lambda$CDM cosmology with massive neutrinos (Hubble parameter $h = 0.67$, baryon density parameter $\Omega_b h^2 = 0.02219$, cold dark matter density parameter $\Omega_c h^2 = 0.12031$, sum of neutrino masses $\Sigma m_{\nu} = 60~{\rm meV}$, amplitude of primordial fluctuations $A_s = 2.151 \times 10^{-9}$, spectral index $n_s = 0.9625$ and optical depth to reionization $~\tau = 0.066$) consistent with measurements from the \Planck satellite \cite{1303.5076,1502.01589,1807.06209}. 

\section{Review of CMB lensing power spectrum estimation}
\label{sec:review}

The lensed CMB is observed in temperature $T(\n)$ and polarization Stokes components $Q(\n)$ and $U(\n)$ as a function of direction on the sky $\n$. The observed CMB sky is lensed by intervening matter such that the lensed sky is related to the unlensed sky through\footnote{We use the flat sky approximation throughout and will eventually work in 2-d Fourier space. Our prescriptions for the cross-only estimator and analytic calculations however all generalize trivially to the full-sky, with simple replacements of curved sky expressions from previous work.} 

\ba
\label{eq:remap}
T(\n)&=&\bar{T}(\n + \balpha) \\
Q(\n)&=&\bar{Q}(\n + \balpha) \nonumber \\
U(\n)&=&\bar{U}(\n + \balpha) \nonumber
\ea
where the bar denotes the unlensed sky and the deflection angle $\balpha$ is related to the integrated line-of-sight gravitational potential out to the surface of last scattering $\phi$ through $\nabla \phi = \balpha$. Our aim is to reconstruct a map $\hat{\phi}(\n)$ of the integrated line-of-sight gravitational potential and then calculate its angular power spectrum $\hat{C}_L^{\phi\phi}$, which can be compared to theoretical models as a function of cosmological parameters.

It is convenient to analyze the polarization fields under the E/B decomposition into $E(\n)$ and $B(\n)$. We use the symbol $X(\n)$ to refer to any of the temperature and polarization fields

$$
X\in\{T,E,B\}.
$$
It is also convenient to work in harmonic space. We denote the 2-d observed CMB (including noise) by $X(\l)$ where $\l$ is the 2-d Fourier space angular wave-number vector. We use the notation:
\be
\int_{\lo+\lt=\L} \big[ \cdots \big] = \int \frac{d^2\lo}{(2\pi)^2} \, \frac{d^2\lt}{(2\pi)^2} \, \big[ \cdots \big] \, (2\pi)^2 \delta^2(\lo+\lt-\L)
\ee

Throughout the paper we will present results in two levels of generality:
the ``isotropic-noise'' case, and the ``anisotropic-noise'' case, which
we now specify in detail as follows.

{\bf In the isotropic-noise case}, we assume that the CMB is observed all-sky
with statistically isotropic noise, and that the observed sky fields $X\ne Y$ have independent noise:
\be
\big\langle X(\l) Y(\l')^* \big\rangle = \big( C_{\ell}^{XY} + N_{\ell}^{XX} \delta_{XY} \big) (2\pi)^2 \delta^2(\l-\l')
\ee
where the angle brackets invoke averaging over realizations of the primary CMB and instrument noise as well as the underlying lensing potential field from large-scale structure. We denote the inverse-variance filtered CMB by $\tX(\l)$:
\be
\tX(\l) = \frac{1}{C_{\ell}^{XX} + N_{\ell}^{XX}} \, X(\l)
\ee
In lens reconstruction estimators (and in optimal $N$-point estimators generally)
the CMB is inverse-variance filtered and the estimator can be written more compactly
in terms of $\tX(\l)$ than $X(\l)$. The isotropic case is useful since it allows us to write down analytic expressions which provide intuition. In some cases like the normalization of the estimator (see below), it also provides a starting point that can be improved with small corrections from Monte Carlo simulations.

{\bf In the anisotropic-noise case}, we make assumptions which are more representative
of real CMB experiments with complex noise models.
In a real experiment, the signal + noise covariance matrix $C+N$ will be non-diagonal
in $\l$. This can be due to the presence of a mask, which for the purposes of this work we consider to be an extreme form of anisotropy where certain regions have infinite noise. More generally, several other sources of non-idealities arise including (a) CMB experiments spending more time observing some regions than others leading to inhomogeneity of the instrument noise (b) atmospheric $1/f$ noise in ground-based experiments leading to possible stripiness or 2-d Fourier-space structure (c) variation of the 2-d Fourier-space structure across the sky due to the scan strategy and sky curvature.   It is usually infeasible to perform the filtering operation $X \rightarrow (C+N)^{-1} X$ exactly.
In this case, $\tX(\l)$ denotes the observed CMB after applying a linear
operation $X \rightarrow \tX$ which approximates the inverse variance filter
$X \rightarrow (C+N)^{-1} X$ as closely as possible (see e.g. \cite{1211.0585,1704.00865,1807.06210,1905.05846,1909.02653}).
We do not assume that an analytic expression is available for either the
filtering operation or the noise model, but we do assume that a black-box
procedure exists for applying the filtering operation $X \rightarrow \tX$
to an arbitrary CMB realization $X$, and that a procedure exists for
making Monte Carlo simulations of the noise. 

\begin{table}

\begin{center}
\begin{tabular}{ll}
$XY$ 		& $f_{XY}({\lo,\lt})$ \vsp
\hline
TT	& $ C_{\ell_1}^{T\nabla T}  \dotfac{1}
	        +   C_{\ell_2}^{T\nabla T}  \dotfac{2}$\vsp
TE	& $ C_{\ell_1}^{TE}\cos 2
		  \varphi_{\lo\lt}
		  \dotfac{1}
    	        +   C_{\ell_2}^{TE}\dotfac{2}$\vsp
TB	& $ C_{\ell_1}^{TE}\sin 2
			\varphi_{\lo\lt}
		  \dotfac{1}
		  $\vsp
EE	        & $[ C_{\ell_1}^{EE}
		  \dotfac{1}
	          + C_{\ell_2}^{EE}
		  \dotfac{2}
			]\cos 2
			\varphi_{\lo\lt}
			$ \vsp
EB		& $ C_{\ell_1}^{EE}
		  \dotfac{1}
		\sin 2
			\varphi_{\lo\lt}
                        $ \vsp
\end{tabular}
\end{center}
\caption{The response of quadratic pairs of CMB modes to lensing \label{tab:response} (See e.g. \cite{1906.08760}).}
\end{table}

\subsection{Quadratic estimator}

The real-space re-mapping in Eq. \ref{eq:remap} translates to Fourier space as a coupling between the $X(\l)$ fields at different wave-numbers that is proportional to the underlying $\phi(\n)$ field at leading order: lensing leads to statistical anisotropy of the CMB when $\phi$ is assumed to be held fixed. This motivates the use of quadratic estimators, weighted sums of pairs of $X(\l)$, to isolate the $\phi(\L)$ field mode by mode\footnote{We use capital $\L$ to denote the Fourier angular wave-numbers of the lensing reconstruction and small-case $\l$ to refer to the same for the CMB maps that are used as inputs in the reconstruction process.}. Physically, in the large-lens limit, CMB lensing reconstructions can be thought of as extracting and stitching together the long-wavelength lensing modes that modulate the small-scale power spectrum of the observed CMB maps. 

For each field pair $XY \in \{ TT, TE, EE, TB, EB \}$ we have a lensing quadratic estimator with kernel $\Fl^{XY}$ formed from inverse-variance filtered fields:
\be
\hphi_{XY}(\L) = A_{XY}(\L) \intlolt \Fl^{XY} \tX(\lo) \tY(\lt)   \label{eq:hphi_def}.
\ee
where $A_{XY}(\L)$ is a normalization for the estimator (see below). For concreteness, we provide the full definition of the kernels here for each quadratic estimator, relating it to notation in \cite{HuOk}.

\be
g_{XY}(\lo,\lt) =
\begin{cases}
 \frac{1}{2}f_{XY}(\lo,\lt),& \text{if $XY=\rm TT,EE$}\\
 f_{XY}(\lo,\lt),& \text{if $XY=\rm EB,TB$}\\
 f_{XY}(\lo,\lt) - f_{XY}(\lt,\lo)\frac{C^{TE}_{\ell_1}}{C^{EE}_{\ell_1}}\frac{C^{TE}_{\ell_2}}{C^{TT}_{\ell_2}},& \text{if $XY=\rm TE$}\\
\end{cases}
\ee
which are related to the lensing response functions
\be
f_{XY}(\lo,\lt) = \frac{\partial }{\partial \phi(\lo+\lt)} \langle X(\lo) Y(\lt) \rangle_{\rm CMB}
\ee
where the angle brackets invoke averaging over realizations of the unlensed CMB. These are shown in Table \ref{tab:response}; note that all spectra that appear there are the lensed spectra, except in the TT estimator, where we use the lensed temperature-gradient power spectrum $C_{\ell}^{T\nabla T}$ (the full non-perturbative response) from \texttt{CAMB}\footnote{\url{https://camb.info/}}\cite{CAMB}: these choices are necessary to reduce the $\Ntwo$ bias in the quadratic estimator to less than a percent \cite{1101.2234,1611.01446,1906.08760} (see also later in this section).  The normalization is then
\be
A_{XY}(\L) = L^2 \left[ \intlolt g_{XY}(\lo,\lt) \frac{1}{C_{\ell}^{XX} + N_{\ell}^{XX}} \frac{1}{C_{\ell}^{YY} + N_{\ell}^{YY}} f_{XY}(\lo,\lt)  \right]^{-1} \label{eq:norm} 
\ee
This normalization results in lensing reconstructions unbiased at the map-level at every position as long as the $\frac{1}{C_{\ell} + N_{\ell}}$ appearing above accurately reflect the filtering applied in $\tX(\l)$, which holds true for the isotropic-noise case. In the anisotropic-nose case (e.g. due to masking and/or optimal filtering), Monte Carlo corrections to the above normalization may be required.  In Eq.~(\ref{eq:hphi_def}) we have defined the quadratic estimator in harmonic space for convenience,
but direct use of the harmonic-space representation would have computational cost $\mathcal{O}(l_{\rm max}^4)$.
In practice, the quadratic estimator is computed using a position-space expression whose
computational cost is the same as a harmonic transform: either $\mathcal{O}(l_{\rm max}^2 \log l_{\rm max})$
in flat sky, or $\mathcal{O}(l_{\rm max}^3)$ on the curved sky. We use the {\tt symlens}\footnote{This Python package is available at \url{https://github.com/simonsobs/symlens/}.} package to symbolically factorize all harmonic-space expressions into sums of position-space products. As a simple example, the TT quadratic estimator reduces to

\ba
\hphi_{TT}(\L) = A_{TT}(\L) \mathcal{F}\left[ {\boldsymbol \nabla} \cdot \left[\tilde{T}(\n) {\boldsymbol \nabla} \tilde{T}_G \right] \right] \label{eq:symlens_tt}
\ea
where ${\boldsymbol \nabla} \tilde{T}_G = \mathcal{F}^{-1} \left[ i\l C_{\ell}^{T\nabla T} \tilde{T}(\l) \right]$ and
\ba
\mathcal{F}\left[ X(\n) \right] &=& \int d^2\n ~ X(\n) e^{-i\l\cdot\n} \nonumber \\
\mathcal{F}^{-1}\left[ X(\l) \right] &=& \int \frac{d^2\l}{(2\pi)^2} ~ X(\l) e^{i\l\cdot\n}
\ea
Position-space expressions on the full sky for the TT, TE, EE, EB and TB estimators can be found in \cite{10.1103/PhysRevD.67.083002} (See also \cite{1807.06210,1908.02016} for minimum-variance estimators that account for the CMB TE correlation). All kernels above are derived by imposing the requirement that the quadratic estimator is unbiased to leading order in $\phi$ and minimum variance, except for the TE estimator, where the requirement for minimum variance is relaxed; an approximation has been made to the kernel that allows it to be factorized into position-space products.

For each quadratic estimator pair $(XY,UV)$, we have a power spectrum estimator
\be
C_L(\hphi_{XY}, \hphi_{UV})
\ee
Since the lensing reconstruction $\hat{\phi}(\n)$ is quadratic in the observed CMB fields $X(\l)$, the process of obtaining the power spectrum $\hat{C}_L^{\phi\phi}$ can be thought of as a special case of trispectrum estimation, i.e. estimation of the four-point function of the CMB. Specifically, we are interested in the primary term of the connected trispectrum, which for a properly normalized quadratic estimator becomes exactly the quantity $C_L^{\phi\phi}$ of interest. However, a naive power spectrum $\hat{C}_L^{\phi\phi} = \langle \hat{\phi}(\L) \hat{\phi}^*(\L) \rangle$ of the lensing reconstruction will also receive contributions from the following sources of bias:
\begin{enumerate}
    \item The Gaussian disconnected component of the four-point function (present even if the CMB maps did not have lensing in them), referred to as the $\mathbf{\Nz}$ bias.
    \item Secondary terms in the connected trispectrum that involve integrals over one factor of $C_L^{\phi\phi}$, and are referred to as $\mathbf{\None}$ bias. The $\None$ bias shows up as excess power ($\sim 10\%$ of $C_L^{\phi\phi}$ at $L\sim 1000$) on small scales \cite{arxiv:0302536,1008.4403}.
    \item A large-scale power suppression $\mathbf{\Ntwo}$ due to higher-order corrections to the quadratic estimate of the lensing potential \cite{1008.4403,1906.08760}, which can be mitigated almost completely by choices described earlier .
    \item The small $\mathbf{\N32}$ bias, which is caused by intrinsic non-Gaussianity in the lensing potential \cite{1605.01392,1806.01157,1806.01216,1906.08760}.
\end{enumerate}
By far the largest of these is the  $\Nz$ bias, which can be significantly larger than the signal of interest $C_L^{\phi\phi}$ itself. It can be thought of as the analog of `noise bias' encountered generally when estimating auto-spectra of a signal, but note that the contributions to this reconstruction noise bias come not just from instrument noise, but also from chance fluctuations of the CMB and foregrounds. It is this large $\Nz$ bias that is the main subject of interest of this work; we aim to make the process of $\hat{C}_L^{\phi\phi}$ reconstruction immune to our assumptions about the instrument noise. In the rest of this section, we will explore in detail how methods so far have attempted to debias the naive lensing power spectrum estimate.

\begin{figure}[t]
    \centering
    \includegraphics[width=0.65\textwidth]{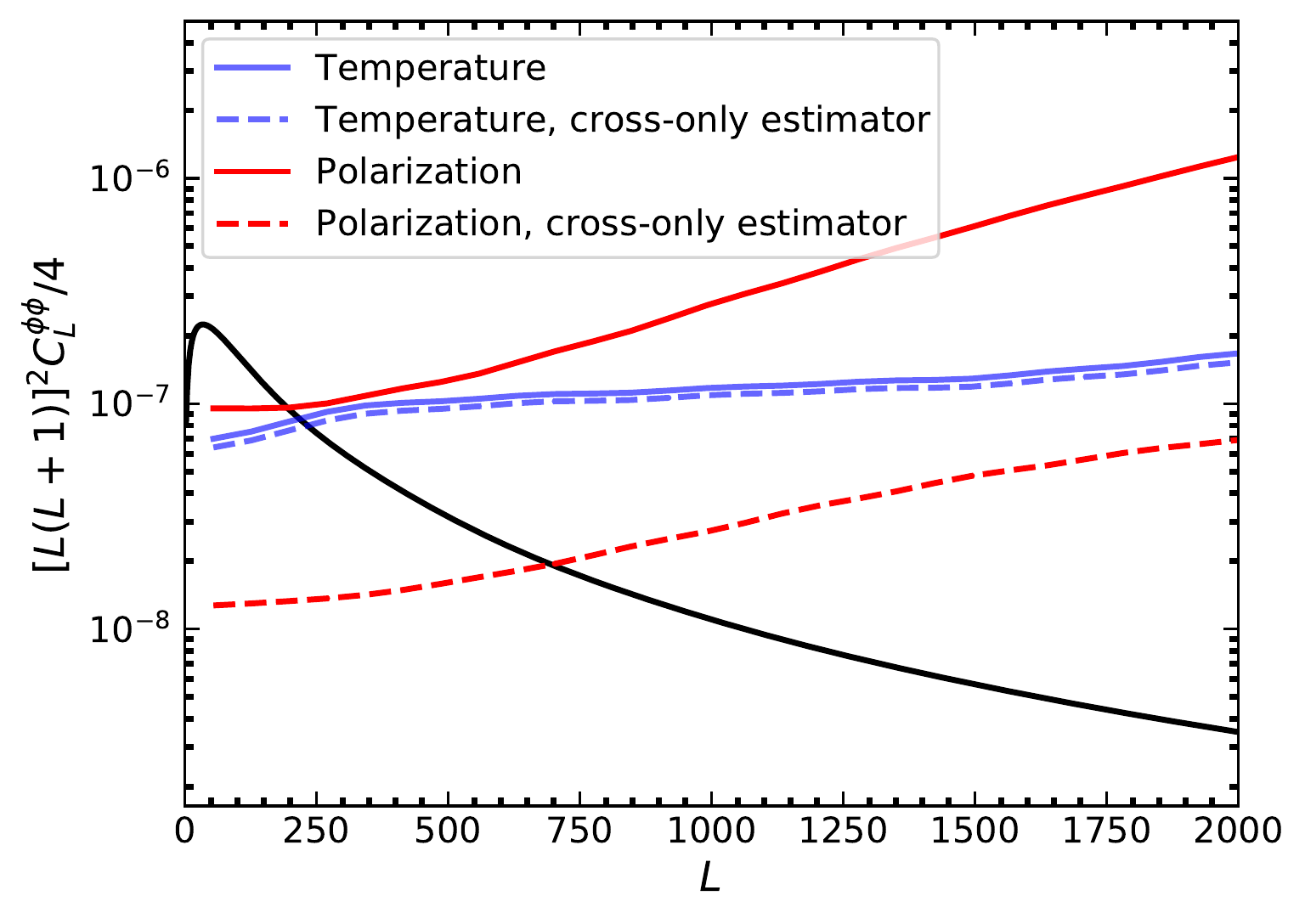}
    \caption{The Gaussian bias to the CMB lensing power spectrum for various estimators that involve CMB temperature data (TTTT estimator in blue) or CMB polarization data (EBEB estimator in red), for a Simons Observatory-like configuration. The bias incurred in a traditional analysis where the full co-added data is used is shown in red and blue solid lines. The smaller Gaussian bias for an estimator that uses our new cross-only estimator is shown in red and blue dashed lines. The true lensing power spectrum is shown in solid black. The cross-only estimator reconstruction noise bias has no dependence on the assumed properties of instrumental or atmospheric noise (which can be difficult to model), and is therefore more robust. Note that this plot shows the {\it reconstruction noise bias} and not the noise per mode; while the noise bias is smaller for the cross-only estimator, the bandpower uncertainty is larger as explored in Section \ref{sec:discussion}, though the difference between the performance of the cross-only and co-add estimator reduces as the number of splits is increased. }
    \label{fig:n0}
\end{figure}

\subsection{$N_0$ bias}

\par\noindent
The $\Nz$-bias $N_{XY,UV}^{(0)}(\L)$ is the expectation value of the lensing power
spectrum estimator that would be obtained if the CMB were an unlensed Gaussian field that had the power spectrum of the observed lensed CMB:
\be
N_{XY,UV}^{(0)}(\L) \equiv \Big\langle C_L(\hphi_{XY}, \hphi_{UV}) \Big\rangle_{\rm Gaussian} \label{eq:N0_nosplit_general}
\ee
{\bf In the anisotropic-noise case}, $\Nz$ can be calculated by Monte Carlo,
from the definitions in Eqs.~(\ref{eq:hphi_def}),~(\ref{eq:N0_nosplit_general}).

\noindent {\bf In the isotropic-noise case}, the $\Nz$ bias can be computed analytically with a single integral. Using Wick's theorem, we get two terms as follows:
\be
N_{XY,UV}^{(0)}(\L) = A_{XY}(\L)A_{UV}(\L)\left[\intlolt \Fl^{XY} \Fmlr^{UV} S_{\lo}^{XU} S_{\lt}^{YV}
   + \intlolt \Fl^{XY} \Fml^{UV} S_{\lo}^{XV} S_{\lt}^{YU}\right]  \label{eq:N0_nosplit_isotropic}
\ee
where the quantity $S_l^{XU}$ is defined for each field pair $(X,U)$ by:
\be
S_{\l}^{XU} = \frac{C_{\ell}^{{\rm obs},XU} + N_{\ell}^{{\rm obs},XU}}{(C_{\ell}^{XX} + N_{\ell}^{XX})(C_{\ell}^{UU} + N_{\ell}^{UU})}  \label{eq:S_def}
\ee
where we take care to distinguish between `observed' power spectra (with the `$\rm obs$' super-script) and power spectra that were used in the filters or kernel of the quadratic estimator. The observed power spectra are the true power spectra that the maps are realizations of: the $\Nz$ bias is proportional to the square of these, and therefore very sensitive to errors made either through incorrectly specifying them in Eq. \ref{eq:S_def} or through incorrectly simulating the noise in the Monte Carlo procedure of Eq. \ref{eq:N0_nosplit_general}. In Figure \ref{fig:n0}, we show in solid lines the $\Nz$ bias estimated using Eq. \ref{eq:S_def} for the TTTT and EBEB estimators for a Simons Observatory-like experiment with 6 $\mu$K-arcmin white noise and a 1.4 arcmin FWHM beam. The bias becomes comparable to the signal at around $L=200$ beyond which it becomes up to three orders of magnitude larger than the signal.

In Eq.~(\ref{eq:N0_nosplit_isotropic}) we have given a harmonic-space representation for the $N_0$-bias,
but in practice the $N_0$-bias is computed using a position-space representation with
lower computational cost. As elsewhere in this article where separable integrals of this form appear, we use the {\tt symlens} package to convert these expressions into position space and evaluate them efficiently.  
The $\Nz$-bias is defined for any pair (XY,UV) of quadratic estimators. We consider all possible combinations in this work but highlight results from the TTTT estimator and the EBEB estimator as examples. 

As a notational point, we have written $S_{\l}^{XU}$ as a function of a 2-d wave-number $\l$
even though it only depends on $\ell=|\l|$, and similarly for the $N_0$ bias in Eq.~(\ref{eq:N0_nosplit_isotropic}).
This is in anticipation of realization-dependent $N_0$ bias, which we discuss next.

\subsection{Realization-dependent $N_0$ bias}

\begin{figure}[t]
    \centering
    \includegraphics[width=0.65\textwidth]{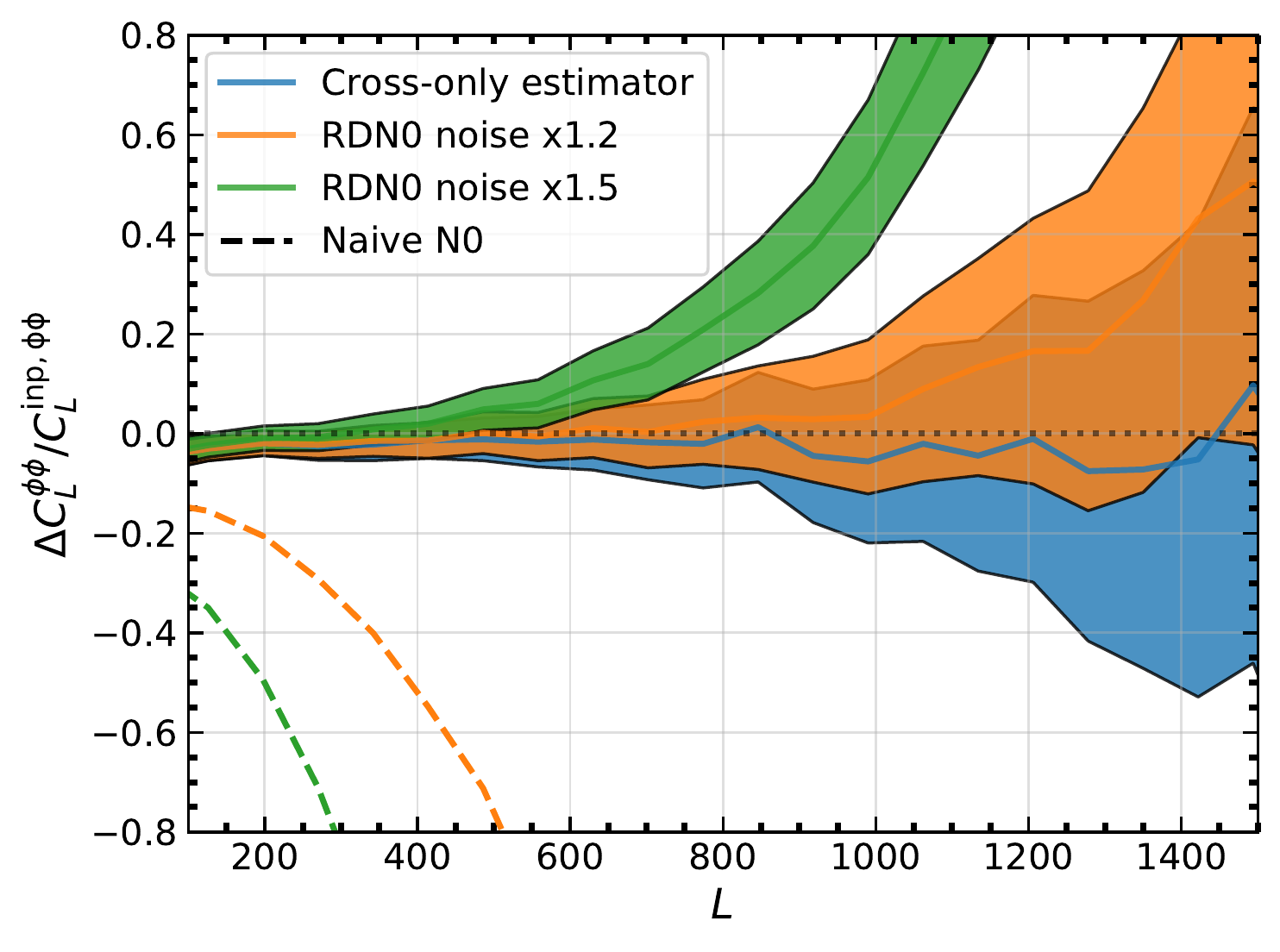}
    \caption{The lensing power spectrum residuals for the EBEB (polarization) estimator after $\Nz$ and $\None$ bias subtraction using various estimators. The dashed lines show the extremely large residuals when the $\Nz$ bias is estimated analytically or from simulations, while mis-estimating the instrumental noise power by 20\% (orange) or 50\% (green). The solid orange and green curves show the residuals for the same mis-estimation when the realization dependent bias subtraction $\RDN0$ from previous work is used, which is relatively more robust, but still shows detectable biases. The residuals from our cross-only estimator (shown here for $m=12$ splits) by construction do not depend on the instrument noise model and are consistent with zero. The mean of the result is therefore unchanged if we assume a different noise level. The error bands show the bandpower $1\sigma$ standard deviation for a Simons Observatory-like configuration with bin width $\Delta L=72$. There is significant overlap between the orange and blue bands at high multipoles. While the $\RDN0$ method works reasonably well even for large (20\%) mismatches of the homogenous power, in a ground-based experiment, it is non-trivial to ensure that the power in every region (due to inhomogeneous noise)  and/or the off-diagonal elements of the covariance matrix match at this level even if the average power over the entire region does. The robustness of the cross-only estimator also becomes important for null tests where much of the CMB sample variance is cancelled.}
    \label{fig:rdn0}
\end{figure}

The simplest debiasing procedure for lens reconstruction is to subtract the simulated or calculated $\Nz$-bias
from the measured $\hphi$ power spectrum.  This has two problems.  First, it is not the optimal
trispectrum estimator for lensing (see e.g \cite{1004.2915,1209.0091}).  Second, the debiasing is ``fragile'', in the sense that if
the assumed covariance matrix of the observed CMB fields has some error $(\Delta C)$, then the bias
to the lensing power spectrum is $\mathcal{O}(\Delta C)$. With naive $\Nz$ subtraction, this is a serious shortcoming that can lead to catastrophic error on the lensing power spectrum when $\Nz$ is much larger than the signal.  As an example, we consider the EBEB estimator bandpowers for a Simons Observatory-like experiment. In Figure \ref{fig:rdn0}, the dashed lines show the relative bias incurred when a mis-estimated naive $\Nz$ is used for debiasing the power spectrum. The result can be approximated as
\be
\frac{\Delta C_L}{C_L} = \frac{C_L+\Nz_{\rm true}-\Nz_{\rm assumed}}{C_L}
\ee
where the noise power spectrum $N_{\ell}^{{\rm obs},XU}$ in Eq. \ref{eq:S_def} differs by 20\% (orange) or 50\% (green) in $\Nz_{\rm assumed}$ relative to its true value in $\Nz_{\rm true}$.  This quickly leads to $\mathcal{O}(1)$ and larger biases for multipoles $L$ greater than a few hundred.

Both of the above-mentioned problems (sub-optimality and catastrophic bias) can be largely mitigated by introducing the realization-dependent $\Nz$ bias $\RDN0$, which can be derived from an Edgeworth expansion of the CMB likelihood \cite{10.1103/PhysRevD.64.083005,0812.1566,1008.4403,1209.0091}.
The $\RDN0$ bias can be written as an expectation value
over Gaussian CMB simulations $X_s$, with the data realization $X_d$ held fixed.
Here, $X \in \{T,E,B\}$ denotes a CMB field as usual. We introduce the notation
\be
\hphi_{XY}^{ab}(\L) \equiv A_{XY}(\L)\intlolt \Fl^{XY} \tX_a(\lo) \tX_b(\lt) \hspace{1cm} (a,b \in \{s,d\})  \label{eq:hphi_ab_def}
\ee
for a ``mixed'' quadratic estimator which can use either data $\tX_d$
or a simulation $\tX_s$ for each of its two inputs.
Then the $\RDN0$ bias can be written (see e.g. \cite{1209.0091,1502.01591}):
\ba
\RDN0_{XY,UV}(\L) &=&
   \Big\langle C_\L\big( \hphi_{XY}^{ds}, \hphi_{UV}^{ds} \big) 
    + C_\L\big( \hphi_{XY}^{ds}, \hphi_{UV}^{sd} \big)
    + C_\L\big( \hphi_{XY}^{sd}, \hphi_{UV}^{ds} \big)
    + C_\L\big( \hphi_{XY}^{sd}, \hphi_{UV}^{sd} \big) \nn \\
&&
    - C_\L\big( \hphi_{XY}^{ss'}, \hphi_{UV}^{ss'} \big)
    - C_\L\big( \hphi_{XY}^{ss'}, \hphi_{UV}^{s's} \big) \Big\rangle_{s,s'}  \label{eq:RDN0_general}
\ea
where $\langle \cdot \rangle_{s,s'}$ denotes an expectation value over two sets of simulations with independent realizations of Gaussian CMB and instrument noise while the data realization $d$ is held fixed.

{\bf In the anisotropic-noise case}, the $\RDN0$-bias can be calculated by Monte Carlo,
from the definitions in Eq.~(\ref{eq:hphi_ab_def}),~(\ref{eq:RDN0_general}).

{\bf In the isotropic-noise case}, the $\RDN0$ bias is given by: 
\ba
\RDN0_{XY,UV}(\L) &=& A_{XY}(\L)A_{UV}(\L)\Big[\intlolt \Fl^{XY} \Fmlr^{UV} \Big( D_{\lo}^{XU} S_{\lt}^{YV} + S_{\lo}^{XU} D_{\lt}^{YV} - S_{\lo}^{XU} S_{\lt}^{YV} \Big) \nn \\
 && \hspace{0.5cm} + \intlolt \Fl^{XY} \Fml^{UV} \Big( D_{\lo}^{XV} S_{\lt}^{YU} + S_{\lo}^{XV} D_{\lt}^{YU} - S_{\lo}^{XV} S_{\lt}^{YU} \Big)\Big]  \label{eq:RDN0_isotropic}
\ea
where the quantity $D_{\l}^{XU}$ is defined for each field pair $(X,U)$ by:
\be
D_{\l}^{XU} \propto \mbox{Re} \big[ \tX_d(\l) \tU_d(\l)^* \big]
\ee
with normalization chosen so that 
$\langle D_{\l}^{XU} \rangle = C_{\l}^{\tX_d \tU_d} = S_{\l}^{XU}$, 
where $S_{\l}^{XU}$ was defined previously in Eq.~(\ref{eq:S_def}).

With this prescription, when the assumed covariance matrix has an error $\Delta C$, the bias to the lensing power spectrum  is $\mathcal{O}(\Delta C^2)$. We explicitly test the robustness of $\RDN0$ subtraction by simulating a lensing bandpower recovery pipeline described in Appendix \ref{sec:sims}. In Figure \ref{fig:rdn0}, the orange and green lines show the relative bias to the lensing power spectrum when the $\RDN0$ bias subtraction in Eq. \ref{eq:RDN0_isotropic} is used but the observed noise power spectra $N_{\ell}^{{\rm obs},XU}$ in Eq. \ref{eq:S_def} is mis-estimated by 20\% and 50\% respectively. The error bands show the $1\sigma$ standard deviation on the bandpowers (with width $\Delta L=72$) for a Simons Observatory-like configuration. While an improvement over the naive $\Nz$ subtraction, a significant bias is still seen at multipoles $L>1000$ in the 20\% mis-estimation case. At 50\% mis-estimation, the $\RDN0$ prescription completely breaks down. While we only explore mis-estimation of the white noise level here, large discrepancies between simulations and data in the off-diagonal two-point function due to noise inhomogeneity are possible, so the 20\% mis-estimation case is relevant for future CMB surveys. These results highlight the need for either accurate simulation of the noise properties or an estimator that is immune to any mis-estimation of the noise. We present such an estimator next.

\section{CMB lensing `cross-only' estimators}
\label{sec:splits}

In the previous section, we reviewed the lensing quadratic estimator $\hphi_{XY}(\L)$,
lensing power spectrum estimator $C_L(\hphi_{XY}, \hphi_{UV})$, reconstruction noise bias
$N^{(0)}_{XY,UV}(\L)$, and realization-dependent reconstruction noise bias $\RDN0_{XY,UV}(\L)$.
These objects are all $N$-point estimators applied to the a CMB map $\tX(\l)$ built
from the full data set.
In this section, we generalize to the case where the data is split into $m$ maps
$\tX^{(i)}(\l)$.

{\bf In the isotropic-noise case}, we assume that the full observed CMB map is the average of $m$ ``split'' maps,
where each split map has independent noise, and isotropic noise power
spectrum $(m N_{\ell}^{XX})$:
\ba
X(\l) &=& \frac{1}{m} \sum_{i=1}^m X^{(i)}(\l) \nn \\
\big\langle X^{(i)}(\l) \, Y^{(j)}(\l')^* \big\rangle
  &=& \Big( C_{\ell}^{XY} + m N_{\ell}^{XX} \delta_{XY} \delta_{ij} \Big) \, (2\pi)^2 \delta^2(\l-\l')
\ea
We propose filtering the splits identically to the co-adds:
\be
\tX^{(i)}(\l) = \frac{1}{C_{\ell}^{XX} + N_{\ell}^{XX}} \, X^{(i)}(\l)
\ee
i.e.~$\tX^{(i)}(\l)$ denotes the $i$-the split map, with inverse-variance filtering
using the full noise power spectrum $N_{\ell}^{XX}$, not the split noise power spectrum
$(m N_{\ell}^{XX})$.  This definition is convenient since the identity
$\tX(\l) = (1/m) \sum_{i=1}^m \tX^{(i)}$ is satisfied.

{\bf In the anisotropic-noise case}, we assume that the full filtered CMB $\tX(\l)$ is the average
of $m$ split maps $\tX^{(i)}(\l)$:
\be
\tX(\l) = \frac{1}{m} \tX^{(i)}(\l)
\ee
with independent noise, but we do not assume that the same filtering operation
$X^{(i)} \rightarrow \tX^{(i)}$ has been applied to each split, or that the noise
model in each split is the same.

\subsection{Cross-only quadratic estimator}

At the map level, the standard quadratic estimator $\hphi_{XY}(\L)$ applied to the full map can be written as a double sum over splits:
\be
\hphi_{XY}(\L) = \frac{1}{m^2} \sum_{ij} \hphi_{XY}^{(ij)}(\L)
\ee
where we have defined
\be
\hphi_{XY}^{(ij)}(\L) = \frac{1}{2} A_{XY}(\L)\intlolt \Fl^{XY} \Big[ \tX^{(i)}(\lo) \tY^{(j)}(\lt) + \tX^{(j)}(\lo) \tY^{(i)}(\lt) \Big]  \label{eq:hphi_split_def}
\ee
Note that we have used a symmetrized definition such that $\hphi_{XY}^{(ij)} = \hphi_{XY}^{(ji)}$.
As in the coadded case, the cross-only estimator $\hphi_{XY}^{ij}$ can be computed efficiently in position space,
using {\tt symlens} to symbolically factorize the harmonic-space expression~(\ref{eq:hphi_split_def})
into sums of position-space products.

At the power spectrum level, the estimator $C_L(\hphi_{XY}, \hphi_{UV})$ can be written as a quadruple sum:
\be
C_L(\hphi_{XY}, \hphi_{UV}) = \frac{1}{m^4} \sum_{ijkl} C_L(\hphi_{XY}^{(ij)}, \hphi_{UV}^{(kl)})
\ee
We define a `cross-only' lensing power spectrum estimator, denoted $C_L^\times(XY,UV)$,
by keeping terms in this quadruple sum in which all four indices $(i,j,k,l)$ are distinct.

Formally, we define:
\be
C_L^\times(XY,UV) = \frac{1}{m(m-1)(m-2)(m-3)} \sum_{ijkl} \gamma_{ijkl} C_L(\hphi^{(ij)}_{XY}, \hphi^{(kl)}_{UV})  \label{eq:Ex_def}
\ee
where the tensor $\gamma_{i_1\cdots i_n}$ is defined for any number of indices $n$ by:
\be
\gamma_{i_1\cdots i_n} = \left\{ \begin{array}{cl}
  1 & \mbox{if $(i_1,\cdots,i_n)$ are all distinct} \\
  0 & \mbox{otherwise}
\end{array} \right. \label{eq:gamma_def}
\ee
This forces the quadruple sum to ignore any contributions to the four-point estimate which repeat a split, e.g. for $m=\{1,2,3,4\}$, $C_L(\hphi^{(12)}_{XY}, \hphi^{(34)}_{UV})$ and $C_L(\hphi^{(31)}_{XY}, \hphi^{(24)}_{UV})$ are allowed (among others) but $C_L(\hphi^{(12)}_{XY}, \hphi^{(14)}_{UV})$ and $C_L(\hphi^{(22)}_{XY}, \hphi^{(34)}_{UV})$ are excluded (among others). For typical CMB experiments, we do not expect significant correlations between the instrument noise in each of $\{T,E,B\}$ on scales relevant for lensing reconstruction, so this estimator is somewhat conservative in the sense that it does not allow repeats of splits even if they come from different elements of $\{T,E,B\}$.

Naively, the cross-only power spectrum estimator $C_L^\times(XY,UV)$ has computational cost $\mathcal{O}(m^4)$ due to the quadruple sum, but as we shall see in the next section it can be done in just $\mathcal{O}(m^2)$.

\subsection{Fast algorithm}

\par\noindent
So far we have defined a quadratic estimator $\hphi^{(ij)}_{XY}$ indexed by a split pair $(i,j)$,
and the coadded quadratic estimator $\hphi_{XY} = (1/m^2) \sum_{ij} \hphi_{XY}^{(ij)}$.
In this section it will be convenient to also define the following quadratic estimators:
\ba
\hphi_{XY}^{(i)} &=& \frac{1}{m} \sum_j \hphi_{XY}^{(ij)} \\
\hphi_{XY}^{\times(i)} &=& \hphi_{XY}^{(i)} - \frac{1}{m} \hphi_{XY}^{(ii)} \\
\hphi_{XY}^{\times} &=& \hphi_{XY} - \frac{1}{m^2} \sum_i \hphi_{XY}^{(ii)}
\ea
Note that the estimators in the second two lines are biased: we have
$\langle \hphi_{XY}^{\times(i)} \rangle = \langle \hphi_{XY}^\times \rangle = (1-1/m) \phi$.

Our $\mathcal{O}(m^2)$ algorithm for computing $C_L^\times(XY,UV)$ is based on the following combinatorial identity:
\be
\gamma_{ijkl} = \gamma_i \gamma_j \gamma_k \gamma_l 
  - \Big[ \delta_{ij} \gamma_k \gamma_l + (\mbox{5 perm.}) \Big]
  + 2 \Big[ \delta_{ijk} \gamma_l + (\mbox{3 perm.}) \Big]
  + \Big[ \delta_{ij} \delta_{kl} + (\mbox{2 perm.}) \Big]
  - 6 \delta_{ijkl}  \label{eq:gamma4}
\ee
where the tensor $\delta_{i_1\cdots i_n}$ is defined for any number of indices $n$ by:
\be
\delta_{i_1\cdots i_n} = \left\{ \begin{array}{cl}
  1 & \mbox{if $i_1 = \cdots = i_n$} \\
  0 & \mbox{otherwise}
\end{array} \right. \label{eq:delta_def}
\ee
The first term allows in all terms, regardless of whether they involve repeats. The second set of terms subtracts cases where two or more splits are repeated. The third set of terms handles cases where three or more splits are repeated. The fourth term handles cases where two splits are repeated twice. The final term subtracts terms where all the splits are identical. Plugging in Eq.~(\ref{eq:gamma4}) on the RHS of Eq.~(\ref{eq:Ex_def}),
the terms which appear can be simplified as follows:
\ba
\sum_{ijkl} \gamma_i \gamma_j \gamma_k \gamma_l C_L(\hphi^{(ij)}_{XY}, \hphi^{(kl)}_{UV})
   &=& m^4 C_L(\hphi_{XY}, \hphi_{UV}) \\
-\sum_{ijkl} \Big[ \delta_{ij} \gamma_k \gamma_l + (\mbox{5 perm.}) \Big] C_L(\hphi^{(ij)}_{XY}, \hphi^{(kl)}_{UV})
   &=& -m^2 \sum_i \Big[ 4 C_L(\hphi_{XY}^{(i)}, \hphi_{UV}^{(i)}) \nn \\
   && \hspace{1cm} + C_L(\hphi_{XY}^{(ii)}, \hphi_{UV}) + C_L(\hphi_{XY}, \hphi_{UV}^{(ii)}) \Big] \\
2 \sum_{ijkl} \Big[ \delta_{ijk} \gamma_l + (\mbox{3 perm.}) \Big] C_L(\hphi^{(ij)}_{XY}, \hphi^{(kl)}_{UV})
   &=& 4 m \sum_i \Big[ C_L(\hphi^{(ii)}_{XY}, \hphi^{(i)}_{UV}) + C_L(\hphi^{(i)}_{XY}, \hphi^{(ii)}_{UV}) \Big] \\
\sum_{ijkl} \Big[ \delta_{ij} \delta_{kl} + (\mbox{2 perm.}) \Big] C_L(\hphi^{(ij)}_{XY}, \hphi^{(kl)}_{UV})
   &=& \sum_{ij} \Big[ C_L(\hphi^{(ii)}_{XY}, \hphi^{(jj)}_{UV}) + 2 C_L(\hphi^{(ij)}_{XY}, \hphi^{(ij)}_{UV}) \Big] \\
-6 \sum_{ijkl} \delta_{ijkl} C_L(\hphi^{(ij)}_{XY}, \hphi^{(kl)}_{UV})
   &=& -6 \sum_i C_L(\hphi^{(ii)}_{XY}, \hphi^{(ii)}_{UV})
\ea

Simplifying, we obtain our final expression for the cross-only lensing power spectrum estimator

\ba
C_L^\times(XY,UV) = \frac{1}{m(m-1)(m-2)(m-3)} 
   \left[ m^4 C_L(\hphi^\times_{XY}, \hphi^\times_{UV}) 
      - 4m^2 \sum_i C_L(\hphi^{(i)\times}_{XY}, \hphi^{(i)\times}_{UV})
      + 4 \sum_{i<j} C_L(\hphi^{(ij)}_{XY}, \hphi^{(ij)}_{UV}) \right] \label{eq:split_estimator}
   \ea
with computational cost $\mathcal{O}(m^2)$.

\subsection{Cross-only $\Nz$-bias}

We define the cross-only $\Nz$-bias by:
\be
\Nzx_{XY,UV}(\L) = \big\langle C_L^\times(XY,UV) \big\rangle_{\rm Gaussian}  \label{eq:N0_split}
\ee
where the expectation value is taken over Gaussian realizations
of the split maps $\tX^{(i)}(\l)$.

{\bf In the anisotropic-noise case}, the cross-only $\Nz$-bias can be
computed by Monte Carlo, from Eq.~(\ref{eq:N0_split}) and the definition
of $C_L^\times(XY,UV)$.

{\bf In the isotropic-noise case}, the cross-only $\Nz$-bias can be computed analytically. After a short calculation using Wick's theorem, the result is:
\be
N_{XY,UV}^{(0)}(\L) = A_{XY}(\L)A_{UV}(\L)\Big[\intlolt \Fl^{XY} \Fmlr^{UV} S_{\lo}^{XU\times} S_{\lt}^{YV\times}
  + \intlolt \Fl^{XY} \Fml^{UV} S_{\lo}^{XV\times} S_{\lt}^{YU\times}\Big]
  \label{eq:n0_split_isotropic}
  \ee
where we have defined:
\be
S_{\l}^{XU\times} = \frac{C_{\ell}^{{\rm obs},XU}}{(C_{\ell}^{XX} + N_{\ell}^{XX})(C_{\ell}^{UU} + N_{\ell}^{UU})}  \label{eq:Sx_def}
\ee
which is the same as $S_{\l}^{XU}$ (Eq.~(\ref{eq:S_def})) except for a missing $N_{\ell}^{{\rm obs}, XU}$ term in the numerator. The missing $N_{\ell}^{{\rm obs}, XU}$ is key; it shows that the cross-only $\Nz$ bias does not care about the actual instrument noise power. In Figure \ref{fig:n0}, we show in dashed lines the $\Nz$ bias for the cross-only estimator for TTTT (blue) and EBEB (red) for a Simons Observatory-like configuration. The level of the bias is greatly reduced for an estimator like EBEB that uses mostly instrument noise dominated scales in polarization. The bias is not completely eliminated because the $\Nz$ bias also receives contributions from CMB fluctuations. We note that it is only the bias that is reduced; the variance of the cross-only estimator is in fact higher, as explored later, although the percentage increase in the variance can be arbitrarily made to approach zero as the number of splits is increased.

\subsection{Cross-only RDN$^{(0)}$ bias}

While the naive cross-only $\Nz$ bias is fully robust to assumptions about the instrument noise, it does not have the partial robustness to assumptions about the CMB and foreground power that the coadd $\RDN0$ estimate has. It is also not optimal. So we explore a realization-dependent cross-only $\Nz$ bias estimate.  In the cross-only case, the $\RDN0$ bias can be represented as an expectation value over
Gaussian simulated split maps $\tX_s^{(i)}(\l)$, in a fixed realization of the
data $\tX^{(i)}_d(\l)$.

We generalize our data/simulation notation from Eq.~(\ref{eq:hphi_ab_def}) for the coadded estimator to allow us to distinguish between splits:
\be
\hphi_{XY}^{ab(ij)}(\L) = A_{XY}(\L)\intlolt \Fl^{XY} \tX_a^{(i)}(\lo) \, \tY_b^{(j)}(\lt)
\hspace{1cm}
(a,b \in \{s,d\} \mbox{ and } i,j \in \{1,\cdots, m\})  \label{eq:hphi_ab_ij}
\ee

The split $\RDN0$ bias is given by the following generalization of Eq.~(\ref{eq:RDN0_general}):
\ba
\RDN0_{XY,UV}(\L) &=& \frac{1}{m(m-1)(m-2)(m-3)} \nn \\
&& 
  \times \sum_{ijkl} \gamma_{ijkl}
   \Big\langle C_\L\big( \hphi_{XY}^{ds(ij)}, \hphi_{UV}^{ds(kl)} \big) 
    + C_\L\big( \hphi_{XY}^{ds(ij)}, \hphi_{UV}^{sd(kl)} \big)
    + C_\L\big( \hphi_{XY}^{sd(ij)}, \hphi_{UV}^{ds(kl)} \big)
    + C_\L\big( \hphi_{XY}^{sd(ij)}, \hphi_{UV}^{sd(kl)} \big) \nn \\
&& \hspace{1.5cm}
    - C_\L\big( \hphi_{XY}^{ss'(ij)}, \hphi_{UV}^{ss'(kl)} \big)
    - C_\L\big( \hphi_{XY}^{ss'(ij)}, \hphi_{UV}^{s's(kl)} \big) \Big\rangle_{s,s'} \nn \\
&=& \Big\langle C^\times_\L\big( \hphi_{XY}^{ds(ij)}, \hphi_{UV}^{ds(kl)} \big) 
    + C^\times_\L\big( \hphi_{XY}^{ds(ij)}, \hphi_{UV}^{sd(kl)} \big)
    + C^\times_\L\big( \hphi_{XY}^{sd(ij)}, \hphi_{UV}^{ds(kl)} \big)
    + C^\times_\L\big( \hphi_{XY}^{sd(ij)}, \hphi_{UV}^{sd(kl)} \big) \nn \\
&& \hspace{1.5cm}
    - C^\times_\L\big( \hphi_{XY}^{ss'(ij)}, \hphi_{UV}^{ss'(kl)} \big)
    - C^\times_\L\big( \hphi_{XY}^{ss'(ij)}, \hphi_{UV}^{s's(kl)} \big) \Big\rangle_{s,s'}
\label{eq:RDN0_split}
\ea
where the efficient algorithm for $C^\times$ in Eq. \ref{eq:split_estimator} is simply applied to each 4-point combination that appeared in Eq.~(\ref{eq:RDN0_general}), and averaged over simulation realizations.

{\bf In the anisotropic-noise case}, the cross-only $\RDN0$-bias can be calculated by Monte Carlo using Eq.~(\ref{eq:RDN0_split}).

{\bf In the isotropic-noise case}, it can be computed analytically. A short calculation using Wick's theorem gives the result, a generalization of Eq.~(\ref{eq:RDN0_isotropic}):
\ba
\RDN0_{XY,UV}(\L) &=& A_{XY}(\L)A_{UV}(\L)\Big[\intlolt \Fl^{XY} \Fmlr^{UV} \Big( D_{\lo}^{XU\times} S_{\lt}^{YV\times} + S_{\lo}^{XU\times} D_{\lt}^{YV\times} - S_{\lo}^{XU\times} S_{\lt}^{YV\times} \Big) \nn \\
 && + \intlolt \Fl^{XY} \Fml^{UV} \Big( D_{\lo}^{XV\times} S_{\lt}^{YU\times} + S_{\lo}^{XV\times} D_{\lt}^{YU\times} - S_{\lo}^{XV\times} S_{\lt}^{YU\times} \Big)\Big] \phantom{XX}\label{eq:RDN0_split_isotropic}
\ea
where $S_{\l}^{XY\times}$ was defined previously in Eq.~(\ref{eq:Sx_def})
and $D_{\l}^{XY\times}$ is defined by the expression
\be
D_{\l}^{XU\times} = 
  \frac{1}{(C_{\ell}^{XX} + N_{\ell}^{XX})(C_{\ell}^{UU} + N_{\ell}^{UU})}
  \left( \frac{1}{m(m-1)}
  \sum_{i\ne j} \mbox{Re} \Big[ X^{(i)}(\l)^* \, U^{(j)}(\l) \Big] \right)
\ee
which involves data two-point power spectra that only use uncorrelated splits.  As in the coadded case, we use {\tt symlens} to convert the Fourier-space
expressions~(\ref{eq:n0_split_isotropic}) and~(\ref{eq:RDN0_split_isotropic})
for $\Nz$ and $\RDN0$ into fast position-space integrals.

As with the naive cross-only $\Nz$, the missing $N_{\ell}^{{\rm obs}, XU}$ shows that cross-only $\RDN0$ bias does not care about the actual instrument noise power by construction. The realization-dependence however makes this estimator additionally robust to errors in the assumed signal power to $\mathcal{O}(\Delta S^2)$ and reduces correlations between bandpowers \cite{1008.4403}. In Figure \ref{fig:rdn0}, we show the residuals for a simulated $m=12$ split analysis for the EBEB estimator in a Simons Observatory-like configuration. We subtract the cross-only $\RDN0$ bias, which does not require us to specify the instrumental noise power. Without having to specify the instrument noise power in the debiasing procedure, we obtain residuals that are consistent with zero given the uncertainty.

\begin{figure}[t]
    \centering
    \includegraphics[width=0.65\columnwidth]{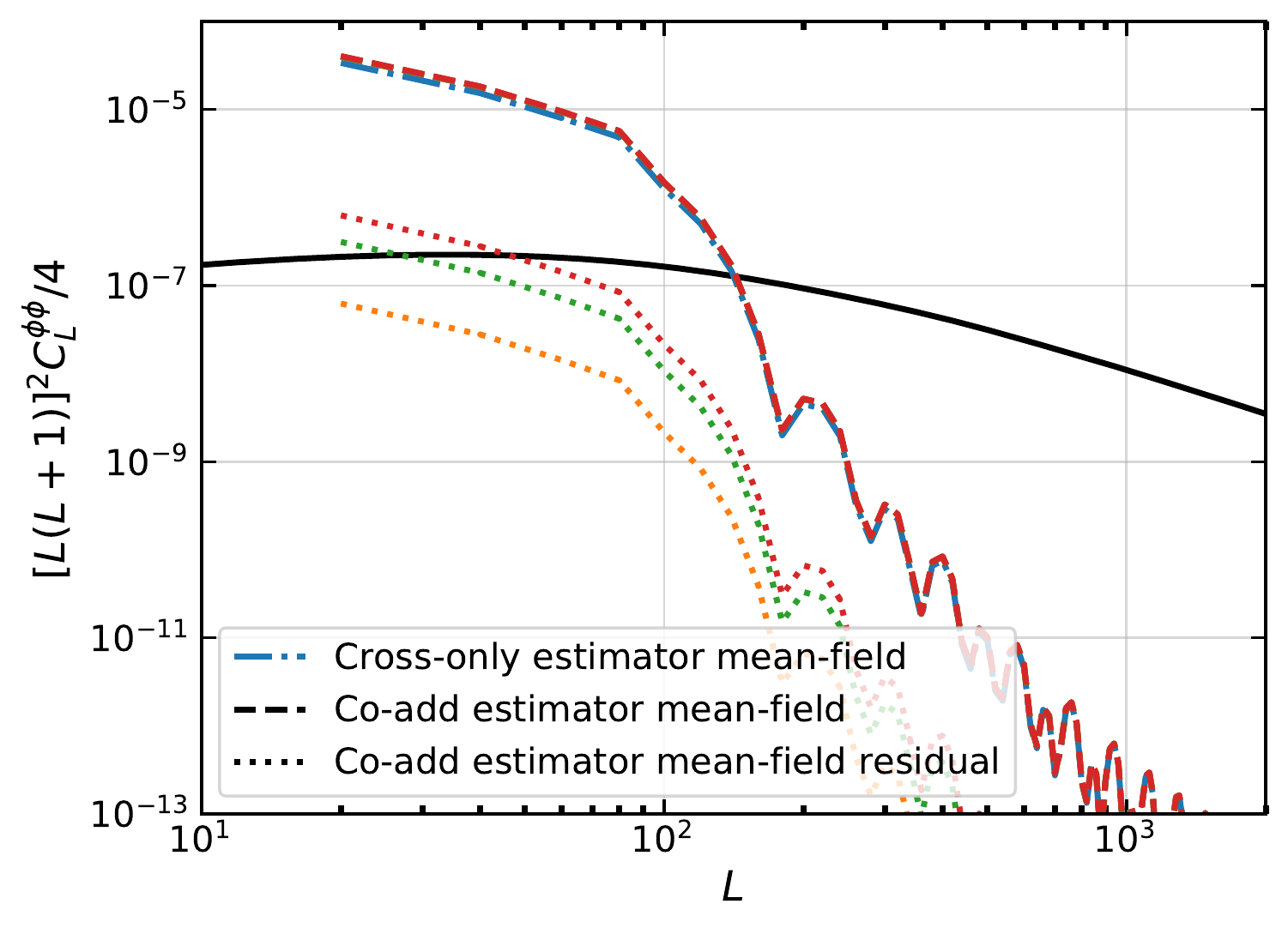}
    \caption{Analytic estimates of the mask-induced mean-field bias compared to the true lensing power spectrum (solid black) for a 1600 sq.deg. patch with a two degree cosine tapered mask. The various dashed lines show the mean-field bias in the traditional co-add estimator when the instrument noise is mis-estimated by 0\% (black dashed), 1\% (orange), 5\% (green) or 10\% (red). These are hard to distinguish on the plot, but the residuals after mean-field subtraction for 1\% (orange), 5\% (green) or 10\% (red) mis-estimation of the instrument noise are shown in dotted lines. These residuals are large and comparable to the lensing signal itself at low multipoles that are important for constraints on primordial non-Gaussianity from sample variance cancellation \cite{1710.09465}. In contrast, the cross-only estimator mean-field (dashed blue) is smaller and has no dependence on the instrumental noise, and is therefore more robust. {\bf The cross-only estimator mean-field residual is by construction zero.}}
    \label{fig:mf}
\end{figure}

\subsection{Mean-field robustness}
\label{sec:mf}
\def\bl{{\boldsymbol{ \ell}}}
\def\dl{d^2{\bl}}
\def\bn{{\boldsymbol{ n}}}

A typical CMB lensing analysis requires parts of the sky to be masked, either due to unobserved regions or due to regions contaminated by foregrounds like Galactic emission. The presence of a mask introduces its own statistical anisotropy which can therefore mimic a lensing signal, resulting in a `mean-field' bias, which rises rapidly on large scales in the lensing reconstruction and depends on the instrument noise power. 

The mean-field bias is typically addressed by averaging Monte Carlo simulations of the lensing reconstruction (where the primary CMB, lensing potential and instrument noise are varied) to obtain a mean-field map $\hphi^{\rm MF} = \langle \hat{\phi} \rangle$, which is then subsequently subtracted from any lensing reconstruction performed on the data. For the co-add estimator, the accuracy of this procedure will depend on the accuracy of the instrument noise assumed in the simulations; since the mean-field power becomes much larger than the signal on the largest lensing scales, this is a serious shortcoming that limits the accuracy with which those scales can be reconstructed. As shown here, our cross-only estimator is once again immune to this.

We focus on the temperature-only (TT) estimator.\footnote{Polarization estimators are more noise-dominated and also add significant weight at the low multipoles of interest. However, they are also smaller in amplitude, so we choose to focus on TT.} Suppose our CMB temperature map is masked by a window $W(\bn)$; denote the Fourier transform of $1-W(\bn)$ as $M(\bl)$. This results in a mean-field bias at the map-level (leading order in $M(\bl)$) \cite{1209.0091}:

$$
\hphi^{\rm MF} = A_{TT}(\L) \int \frac{d^2\bl_1}{(2\pi)^2} \frac{f_{TT}(\lo,\lt)f_{M}(\lo,\lt)}{2(C^{TT}_{\ell_1}+N^{TT}_{\ell_1})(C^{TT}_{\ell_2}+N^{TT}_{\ell_2})} M_{\boldsymbol{L}}
$$
where $f_{TT}(\lo,\lt)$ is the usual lensing response for the TT estimator only involving sky spectra (from Table \ref{tab:response}) and
$$
f_{M}(\lo,\lt) = -(C^{{\rm obs},TT}_{\ell_1}+N^{{\rm obs},TT}_{\ell_1}) - (C^{{\rm obs},TT}_{\ell_2}+N^{{\rm obs},TT}_{\ell_2})
$$
for the co-added estimator, whereas for the cross-only estimator
$$
f_{M}(\lo,\lt) = -C^{{\rm obs},TT}_{\ell_1} - C^{{\rm obs},TT}_{\ell_2}
$$
The mean-field power spectrum is then:

$$
C^{\rm MF}_L = \left(A_{TT}(\L) \int \frac{d^2\bl_1}{(2\pi)^2} \frac{f_{TT}(\lo,\lt)f_{M}(\lo,\lt)}{2(C^{TT}_{\ell_1}+N^{TT}_{\ell_1})(C^{TT}_{\ell_2}+N^{TT}_{\ell_2})} \right)^2 C^{\rm mask}_L
$$
where $C^{\rm mask}_L = \langle M_{\boldsymbol{L}}
M_{\boldsymbol{L}}^{*} \rangle$

Since the mean-field is obtained from Monte Carlo 
simulations, the spectrum $N^{{\rm obs},TT}_{\ell}$ that appears in the numerator can be
perturbed to infer the cost of mis-specifying the noise model. For the cross-only
estimator, $N^{{\rm obs},TT}_{\ell}$ does not appear in the mean-field, and so the mean-field is both reduced in size and
robust to mis-specification of the noise model.  In Figure \ref{fig:mf}, we show how the dependence of the mean-field on the instrument noise can lead to large residuals on large angular scales in the lensing reconstruction. The cross-only estimator on the other hand has a mean-field that is robust against mis-simulation of the instrument noise.  The scales affected ($L<100$) are important for constraints on primordial non-Gaussianity through sample variance cancellation \cite{1710.09465}. We have assumed the flat-sky approximation and isotropic filtering here. While the effect on the co-add estimator may be quantitatively different for a realistic curved-sky analysis that uses optimal filtering, the robustness of the cross-only estimator does not depend on these details. It should be noted that instrumental systematics can also affect the signal in the CMB maps, which can cause residual mean-fields. The cross-only estimators presented here are not just useful for the baseline bandpowers used in a cosmological analysis, but also for various investigative null-tests. For example, when investigating the effect of systematics like inaccurate anisotropic beams in the simulations, it is useful to calculate the debiased lensing power spectrum of differences of maps from detector sets or seasons. In such a difference, much of the CMB sample variance cancels allowing for a null test bandpower uncertainty that is significantly smaller. In order to isolate biases that affect the signal in the CMB map, it therefore is important to use the cross-only estimator we present here so as to disentangle the effects of noise mis-simulation.

\subsection{A partially robust two-split estimator}
As presented in \cite{1807.06210}, in situations where only $m=2$ splits are available, a simpler estimator that is only partially robust can be built:
\be
C_L^{\rm two-split}(XY,UV) = \frac{1}{2}\left[ C_L(\hphi^{(11)}_{XY}, \hphi^{(22)}_{UV}) + C_L(\hphi^{(22)}_{XY}, \hphi^{(11)}_{UV})\right].
\ee 
The $\Nz$ bias is then once again given by the expressions in Eq. \ref{eq:n0_split_isotropic} and Eq. \ref{eq:RDN0_split_isotropic}, and therefore remains robust to mis-simulation of the noise. In addition, since each split is only repeated twice, the instrument noise enters only through two-point functions; hence, this estimator is also insensitive to non-Gaussianity in the noise.  However, the mean-field bias in the $m=2$ estimator will still depend on the instrument noise. The combination $C_L(\hphi^{(12)}_{XY}, \hphi^{(12)}_{UV}) $ on the other hand has no dependence of the mean-field bias on instrument noise, but is sensitive to Gaussian and non-Gaussian contributions from the instrument noise \cite{1807.06210}. Robustness against all these sources requires the more general $m\geq4$ cross-only estimator presented in this work.

\section{Discussion}
\label{sec:discussion}

We have provided a new estimator for the CMB lensing power spectrum that is by construction immune to any assumptions made in modeling or simulating the instrument noise. The estimator requires at least four splits of the CMB map that have independent realizations of the noise, and is constructed by excluding all terms in the four-point lensing power estimator that repeat a split. In such an estimator, no property of the instrument noise will affect the mean of the estimator, so even unknowns such as potential non-Gaussianity in the noise will not bias CMB lensing estimation.  In particular, the power-spectrum estimation typically requires a large $\Nz$ bias to be estimated from Monte Carlo simulations and subtracted; our estimator eliminates any dependence of this procedure on the instrument noise. In addition, measurements of the power spectrum on very large scales ($L<10-100$) are typically limited by the ability of simulations to accurately estimate the mask-induced `mean-field' bias; this procedure is also made robust against mis-simulation of instrument noise by our estimator. We have also provided a fast algorithm for the cross-only estimator that scales with the number of splits $m$ as $\mathcal{O}(m^2)$ as opposed to a naive $\mathcal{O}(m^4)$ implementation.

\subsection{Performance}
This robustness is of course achieved by discarding some of the data, so the question naturally arises whether the lensing signal-to-noise ratio is significantly affected. For a generic $N$-point estimator, the ratio of variance of a cross-only estimator with $m$ splits relative to a co-add estimator can be estimated in the noise-dominated limit from the reciprocal of the ratio of the number of terms in a co-add estimator $m^N$ (each of $N$ points can be any of $m$ splits) over the number of terms in a cross-only estimator $m!/(m-N)!$  (the number of permutations of $m$ splits when $N$ are chosen) 

$$
\frac{m^N(m-N)!}{m!}
$$
In this limit, it would appear that 4-point lensing estimators are significantly more affected by splitting relative to 2-point CMB power spectrum estimation, with e.g. 10.67 times higher noise variance for $m=4$ in the 4-point case compared to 1.33 for the 2-point case. However, CMB lensing estimators typically restrict themselves to using multipoles $\ell<3000$ to avoid foreground bias, and in this multipole range measurements are increasingly becoming signal dominated in temperature and E-mode polarization. We therefore expect that in practice the loss in signal-to-noise should not be substantial.

\begin{figure}[t]
    \centering
    \includegraphics[width=0.65\textwidth]{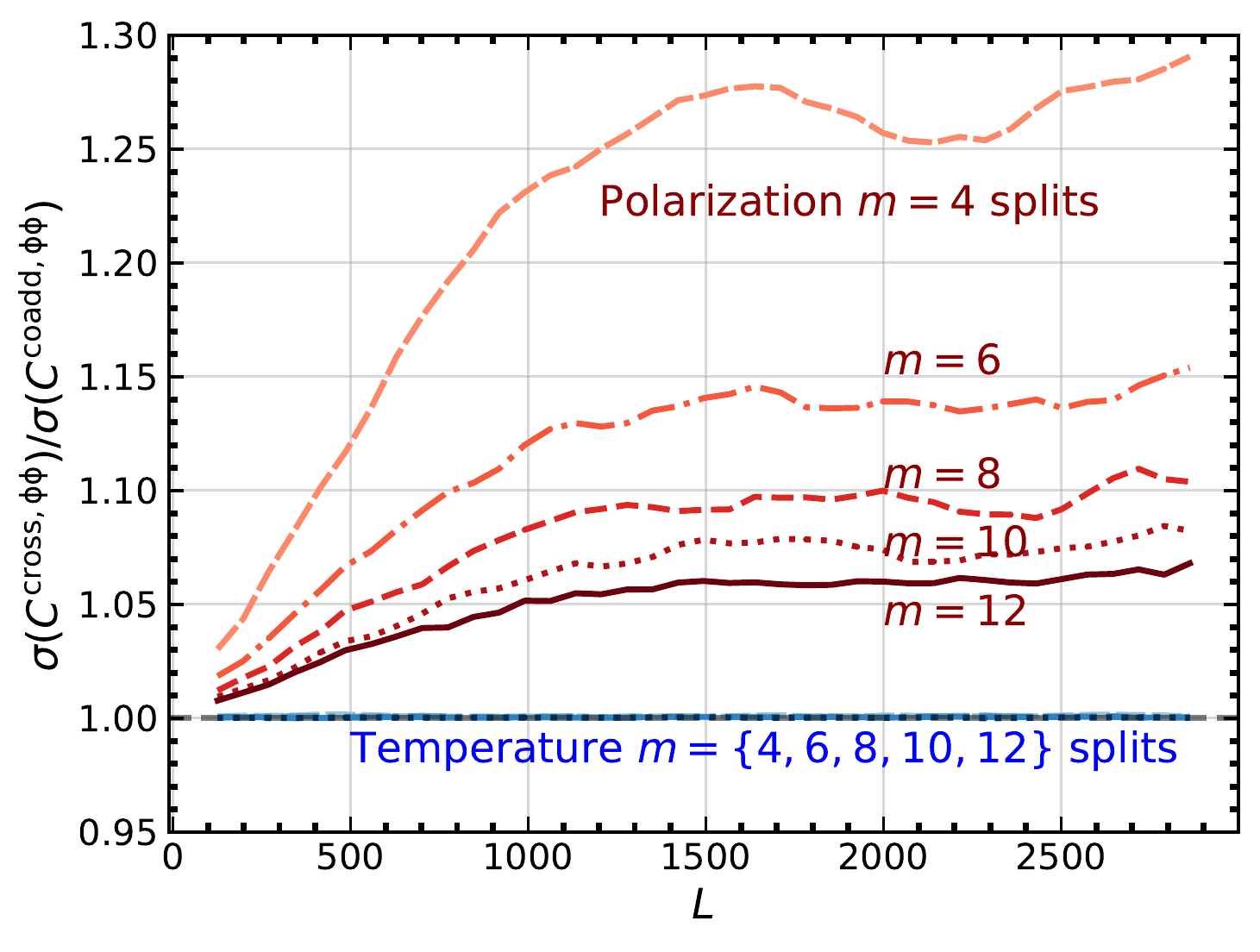}
    \caption{The ratio of bandpower standard deviations for the cross-only estimator with respect to the traditional co-add estimator, calculated from simulations of a Simons Observatory-like configuration. The TTTT temperature estimator (blue curves of various line styles; not distinguishable in this plot) incurs no signal-to-noise ratio penalty in this configuration even with the minimum number of splits $m=4$ because temperature measurements below $\ell=3000$ are primarily signal dominated. The EBEB polarization estimator incurs a penalty of up to 23\% for $m=4$ splits and $L<1000$, but this is reduced to close to 5\% for $m=12$ splits and $L<1000$. With the $\mathcal{O}(m^2)$ scaling of the computational cost, it should be noted that the $m=12$ estimator would take 9 times longer to compute than the $m=4$ estimator. The overall weight of the EBEB estimator in a final minimum-variance combination of all temperature+polarization estimators should also be taken into account when considering these trade-offs.}
    \label{fig:error}
\end{figure}

To check this intuition, we simulate lensing bandpower recovery for various experiments as described in Appendix \ref{sec:sims}. After the appropriate $\RDN0$ and $\None$ subtraction, we compare the ratio of the bandpower standard deviation as a function of lensing reconstruction multipole $L$ and number of splits $m$. Our results for the TTTT and EBEB estimators in a Simons Observatory-like configuration are shown in Figure \ref{fig:error}. For the TTTT estimator, we do not detect any degradation in signal-to-noise, a consequence of the fact that the temperature field measurement from Simons Observatory is mostly signal-dominated out to $\ell<3000$. For the EBEB estimator with the minimum number of splits $m=4$, we find a 5-30\% increase in the bandpower uncertainty depending on multipole. However, this is reduced to a 1-5\% increase in uncertainty if $m=12$ splits are used. In practice though, the EBEB estimator contributes low weight in a minimum-variance combination of all the estimator power spectra, and so the overall penalty on signal-to-noise ratio for the full temperature+polarization measurement will be much smaller. This consideration should be squared with the fact that an $m=12$ analysis will on average take nine times longer than an $m=4$ analysis. For other temperature-polarization estimator combinations (e.g. TTTE) and other experiments considered (see Appendix \ref{sec:sims}), we find that these general conclusions hold:

\begin{enumerate}
\item The signal-to-noise loss from using the cross-only estimator is not substantial even for $m=4$ splits.
\item The loss depends on how signal-dominated the constituent fields of the four-point estimator are.
\item It approaches zero as the number of splits is increased.
\end{enumerate}

There are in practice limits to how finely one can split observations. The requirement that the splits should have independent noise means that splits should not be interleaved on time-scales shorter than several hours given the noise properties of typical CMB surveys. This results in each split consisting of chunks of data at least several hours long, which limits the total number of such chunks one can allocate to the different splits. If the number of splits gets too high, then there will be too few chunks of data to go around, leading to some splits having uneven or even incomplete coverage of the survey area. Uneven coverage makes it difficult to build an optimal $\tilde X^{(i)}$ by making the noise properties strongly position-dependent, while incomplete coverage means that the effective number of splits in those parts of the sky is lower than what was requested. All else equal, the maximum practical number of splits is determined by the size of the survey area. Wide surveys, which spread their data out over a large area, can afford fewer splits than those that focus the same amount of data in a small area. That said, as Figure~\ref{fig:rdn0} shows, eight splits is enough to go below a 10\% loss in S/N even for the worst case of a strongly noise-dominated estimator. This number is easily achievable even for the ongoing wide-area surveys of ACT (provided several years of data are analyzed together), Simons Observatory and CMB-S4.

\subsection{Other biases}
\label{sec:higher} 

Our cross-only estimator does not require any modification to the usual procedures for mitigating the $\None$ and $\Ntwo$ bias. We have numerically verified this by obtaining unbiased spectra after applying the same mitigation techniques to both the co-add estimator and the cross-only estimator, i.e. the use of non-perturbative responses and lensed power spectra for the $\Ntwo$ bias and the subtraction of the same Monte Carlo $\None$ bias calculated from simulations of the co-add. Physically, this is expected, since both of these biases arise from higher-order corrections proportional to the lensing potential and have no dependence on the instrumental noise.

\subsection{Future work}

The ideas presented here apply to general $N$-point estimation, e.g. primordial non-Gaussianity bispectra and trispectra estimation. Such analyses can also be made robust to assumptions about the instrument noise using cross-only combinations, can be done efficiently with better than $\mathcal{O}(m^N)$ algorithmic complexity, and incur lower signal-to-noise penalties than naively expected due to signal-dominated scales in the relevant fields. We leave detailed exploration of extensions to primordial non-Gaussianity and $N$-point estimation to future work.

The quadratic estimator can be thought of as the first step in a Newton-Raphson iteration to find the maximum likelihood $\hat{\phi}(\n)$: as such, it can be sub-optimal in the high signal-to-noise regime (typically instrument noise levels well below 5 $\mu$K-arcmin / Simons Observatory white noise levels) where iterative or Bayesian techniques will be required \cite{2015ApJ...808..152A,2017PhRvD..96f3510C,2019PhRvD.100b3509M,2020arXiv200200965M}. The application of split-based methods in this regime possibly requires non-trivial modifications to our prescriptions, and future work investigating this will be required for robust lensing power spectra from CMB-S4 and beyond. Nevertheless, we have laid out a prescription that greatly relaxes instrument simulation requirements for the next generation of CMB experiments while still allowing CMB lensing to deliver accurate cosmological information.

\acknowledgments
We thank the anonymous referee for valuable suggestions that helped improve this paper. This research made use of Astropy\footnote{http://www.astropy.org}, a community-developed core Python package for Astronomy \citep{astropy:2013, astropy:2018}. We also acknowledge use of the \texttt{matplotlib}~\cite{Hunter:2007} and \texttt{numpy} ~\cite{numpy} packages and use of the Boltzmann code \texttt{CAMB}~\cite{CAMB} for calculating theory spectra. Research at Perimeter Institute is supported
in part by the Government of Canada through
the Department of Innovation, Science and Industry Canada and by the Province of Ontario through the Ministry of Colleges and Universities. BDS acknowledges support from a European Research Council (ERC) Starting Grant under the European Union’s Horizon 2020 research and innovation programme (Grant agreement No. 851274) and from an STFC Ernest Rutherford Fellowship. Flatiron Institute is supported by the Simons Foundation.

\bibliography{msm}

\appendix

\section{Simulations}
\label{sec:sims}

We compare the performance and accuracy of the co-add estimator with the cross-only estimator by simulating a lensing bandpower reconstruction pipeline. Our simulations are generated under the flat-sky approximation, are periodic and unmasked, and add homogeneous white noise. This allows us to use and validate the isotropic-case analytic expressions from Section \ref{sec:review} and \ref{sec:splits}.

We start by generating periodic Gaussian random field realizations given the unlensed primary CMB power in temperature and polarization and the lensing potential power evaluated using \texttt{CAMB} at our fiducial cosmology. These are generated on $20 ~\rmn{deg} \times 20 ~\rmn{deg}$ patches with 1.5 arcminute wide pixels. These choices are motivated by our requirement that the lensing be accurate out to multipoles of around $\ell=3000$ and that reliable reconstructions can be obtained down to $L=100$\footnote{We leave detailed verification down to larger scales with full-sky estimators to future work, but we do not expect differences in the qualitative conclusions.}.  The Gaussian CMB I/Q/U components are then lensed using 5th-order spline interpolation\footnote{This is done using the routines in {\tt pixell} at \url{https://github.com/simonsobs/pixell/}} with a deflection field $\vec{\alpha}$ obtained from the simulated Gaussian lensing potential field using $\nabla \phi = \vec{\alpha}$. The lensed patches are convolved with a Gaussian beam of full-width-half-maximum (FWHM) of 1.4 arcminutes, which roughly corresponds to the effective beam at the dominant 150 GHz channel for experiments like ACT, SPT, Simons Observatory and CMB-S4.

We then simulate various analyses where $m$ splits of the data are available, with $m=4,6,8,10,12$. For each of these, $m$ realizations of white noise Gaussian random fields are added as instrumental noise. Our baseline analysis (shown in the Figures in this article) uses a white noise root-mean-square (RMS) level of 6 $\mu$K-arcmin, corresponding roughly to Simons Observatory data (without atmospheric noise), but we also consider variations with RMS noise of 3 $\mu$K-arcmin, 10 $\mu$K-arcmin and 30 $\mu$K-arcmin.

For each of the $m$-split cases and experiments under consideration, we proceed with the CMB lensing bandpower analysis either on the mean of the $k$ splits (which we call the ``coadd estimator'' and corresponds to a traditional analysis) or through the ``cross-only'' estimator of Eq. \ref{eq:split_estimator}. We use CMB multipoles in the range $100 \leq \ell \leq 3000$ in the reconstruction. We subtract the isotropic $\RDN0$ for each case (Eq. \ref{eq:RDN0_isotropic} for the co-add estimator and Eq. \ref{eq:RDN0_split_isotropic} for the cross-only estimator). We subtract the same $\None$ bias from both the split and co-add estimator: this bias is calculated using Monte Carlo simulations of the co-add estimator as prescribed in \cite{1412.4760,1611.09753} and constructed from 3333 realizations. For the TTTT and EBEB combinations in the Simons Observatory-like configuration shown in the Figures in this paper, we use 8000 realizations of the bias-subtracted bandpowers to estimate their scatter. We also estimate the bias-subtracted bandpower scatter for 400 realizations of all other estimator combinations (e.g. TTTE) for each of 3, 6, 10 and 30 $\mu$K-arcmin white noise.

\end{document}